\documentclass{article}
\usepackage{arxiv}
\usepackage[utf8]{inputenc}
\usepackage[T1]{fontenc}
\usepackage[inline]{enumitem}
\usepackage{booktabs}
\usepackage{amsmath}
\usepackage{amsfonts}
\usepackage{amssymb}
\usepackage{enumitem}
\usepackage{mathbbol}
\usepackage{mathtools}
\usepackage{hyperref}
\usepackage{url}
\usepackage{nicefrac}
\usepackage{microtype}
\usepackage{natbib}
\usepackage{multirow}
\usepackage{subcaption}

\newcommand{\eg}{\emph{e.g.}}
\newcommand{\wrt}{\emph{w.r.t. }}

\newcommand{\vs}{\emph{vs. }}

\title{Conformer-Kernel with Query Term Independence for Document Retrieval}

\author{
  Bhaskar Mitra \\
  Microsoft, UCL \\
  \texttt{bmitra@microsoft.com} \\
   \And
  Sebastian Hofst\"{a}tter \\
  TU Wien \\
  \texttt{s.hofstatter@tuwien.ac.at}
   \And
  Hamed Zamani and Nick Craswell \\
  Microsoft \\
  \{\texttt{hazamani, nickcr}\}\texttt{@microsoft.com}
}

\begin{document}
\maketitle

\begin{abstract}
\end{abstract}
The Transformer-Kernel (TK) model has demonstrated strong reranking performance on the TREC Deep Learning benchmark---and can be considered to be an efficient (but slightly less effective) alternative to BERT-based ranking models.
In this work, we extend the TK architecture to the full retrieval setting by incorporating the query term independence assumption.
Furthermore, to reduce the memory complexity of the Transformer layers with respect to the input sequence length, we propose a new Conformer layer.
We show that the Conformer's GPU memory requirement scales linearly with input sequence length, making it a more viable option when ranking long documents.
Finally, we demonstrate that incorporating explicit term matching signal into the model can be particularly useful in the full retrieval setting.
We present preliminary results from our work in this paper.


\section{Introduction}
\label{sec:intro}
In the inaugural year of the TREC Deep Learning track~\citep{craswell2019overview}, Transformer-based~\citep{vaswani2017attention} ranking models demonstrated substantial improvements over traditional information retrieval (IR) methods.
Several of these approaches---\eg, \citep{yilmaz2019h2oloo, yan2019idst}---employ BERT~\citep{devlin2018bert}, with large-scale pretraining, as their core architecture.
Diverging from the trend of BERT-scale models, \citet{Hofstaetter2020_ecai} propose the Transformer-Kernel (TK) model with few key distinctions:
\begin{enumerate*}[label=(\roman*)]
    \item TK uses a shallower model with only two Transformer layers,
    \item The parameters of the model are randomly initialized prior to training (skipping the computation-intensive pretraining step), and finally
    \item TK encodes the query and the document independently of each other allowing for offline precomputations for faster response times.
\end{enumerate*}
Consequently, TK achieves competitive performance at a fraction of the training and inference cost of its BERT-based peers.

Notwithstanding these efficiency gains, the TK model shares two critical drawbacks with other Transformer-based models. Firstly, the memory complexity of the self-attention layers is quadratic $\mathcal{O}(n^2)$ with respect to the length $n$ of the input sequence.
This restricts the number of document terms that the model can inspect under fixed GPU memory budget.
A trivial workaround involves inspecting only the first $k$ terms of the document.
This approach can not only negatively impact retrieval quality, but has been shown to specifically under-retrieve longer documents~\citep{hofstatter2020improving}.
Secondly, in any real IR system, it is impractical to evaluate every document in the collection for every query---and therefore these systems typically either enforce some sparsity property to drastically narrow down the set of documents that should be evaluated or find ways to prioritize the candidates for evaluation.
TK employs a nonlinear matching function over query-document pairs.
Therefore, it is not obvious how the TK function can be directly used to retrieve from the full collection without exhaustively comparing every document to the query.
This restricts TK's scope of application to late stage reranking of smaller candidate sets that may have been identified by simpler retrieval models.

In this work, we extend the TK architecture to enable direct retrieval from the full collection of documents.
Towards that goal, we incorporate three specific changes:
\begin{enumerate}
    \item To scale to long document text, we replace each instance of the Transformer layer with a novel Conformer layer whose memory complexity is $\mathcal{O}(n \times d_\text{key})$, instead of $\mathcal{O}(n^2)$.
    \item To enable fast retrieval with TK, we incorporate the query term independence assumption~\citep{mitra2019incorporating} into the architecture.
    \item And finally, as \citet{mitra2016desm, mitra2017learning} point out, lexical term matching can complement latent matching models, and the combination can be particularly effective when retrieving from the full collection of candidates.
    So, we extend TK with an explicit term matching submodel to minimize the impact of false positive matches in the latent space.
\end{enumerate}

We describe the full model and present preliminary results from our work in this paper.
\section{Related work}
\label{sec:related}

\subsection{Scaling self-attention to long text}
\label{sec:related-long}

The self-attention layer, as proposed by \citet{vaswani2017attention}, can be described as follows:

\begin{align}
    \text{Self-Attention}(Q, K, V) &= \Phi(\frac{QK^\intercal}{\sqrt{d_k}}) \cdot V
\end{align}
Where, $Q \in \mathbb{R}^{n \times d_\text{key}}$, $K \in \mathbb{R}^{n \times d_\text{key}}$, and $V \in \mathbb{R}^{n \times d_\text{value}}$ are the query, key, and value matrices---and $d_\text{key}$ and $d_\text{value}$ are the dimensions of the key and value embeddings, respectively, and $n$ is the length of the input sequence.
We use $\Phi$ to denote the softmax operation applied along the last dimension of the input matrix.

The quadratic $\mathcal{O}(n^2)$ memory complexity of self-attention is a direct consequence of the component $QK^\intercal$ that produces a matrix of size $n \times n$.
Recently, an increasing number of different approaches have been proposed in the literature to get around this quadratic complexity.
Broadly speaking, most of these approaches can be classified as either:
\begin{enumerate*}[label=(\roman*)]
    \item Restricting self-attention to smaller windows over the input sequence which results in a memory complexity of $\mathcal{O}(n \times m)$, where $m$ is the window size---\eg,~\citep{parmar2018image, dai2019transformer, yang2019xlnet, sukhbaatar2019adaptive}, or
    \item Operating under the assumption that the attention matrix is low rank $r$ and hence finding alternatives to explicitly computing the $QK^\intercal$ matrix to achieve a complexity of $\mathcal{O}(n \times r)$---\eg,~\citep{kitaev2019reformer, roy2020efficient, tay2020sparse, wang2020linformer}, or
    \item A hybrid of both approaches---\eg,~\citep{child2019sparse, beltagy2020longformer, Wu2020Lite}.
\end{enumerate*}
In the IR literature, recently \citet{hofstatter2020improving} have extended the TK model to longer text using the local window-based attention approach.
Other more general approaches to reducing the memory footprint of very deep models, such as model parallelization have also been extended to Transformer models~\citep{shoeybi2019megatron}.
For more general primer on self-attention and Transformer architectures, we point the reader to \citet{weng2020attention, weng2020transformer}.


\subsection{Full retrieval with deep models}
\label{sec:related-retrieval}

Efficient retrieval using complex machine learned relevance functions is an important challenge in neural IR~\citep{mitra2018introduction,Guo:etal:IPM2019}.
One family of approaches involves the dual encoder architecture where the query and document are encoded independently of each other, and efficient retrieval is achieved using approximate nearest-neighbour search~\citep{lee2019latent, chang2020pre, karpukhin2020dense, ahmad2019reqa, khattab2020colbert} or by employing other data structures, such as learning an inverted index based on latent representations~\citep{zamani2018neural2}.
Precise matching of terms or concepts may be difficult using query-independent latent document representations~\citep{luan2020sparse}, and therefore these models are often combined with explicit term matching methods~\citep{nalisnick2016improving, mitra2017learning}.
\citet{xiong2020approximate} have recently demonstrated that the training data distribution can also significantly influence the performance of dual encoder models under the full retrieval setting.
Auxilliary optimization objectives can also help guide the training of latent matching models to find solutions that emphasize more precise matching of terms and concepts~\citep{rosset2019axiomatic}.

An alternative approach assumes query term independence (QTI) in the design of the neural ranking model~\citep{mitra2019incorporating}.
For these family of models, the estimated relevance score $S_{q,d}$ is factorized as a sum of the estimated relevance of the document to each individual query term.

\begin{align}
    S_{q,d} &= \sum_{t \in q}{s_{t,d}}
\end{align}

Readers should note that the QTI assumption is already baked into several classical IR models, like BM25~\citep{robertson2009probabilistic}.
Relevance models with QTI assumption can be used to precompute all term-document scores offline.
The precomputed scores can be subsequently leveraged for efficient search using inverted-index data structures.

Several recent neural IR models~\citep{mitra2019incorporating, dai2019evaluation, dai2019deeper, mackenzie2020efficiency, daicontext, macavaney2020expansion} that incorporate the QTI assumption have obtained promising results under the full retrieval setting.
Document expansion based methods~\citep{nogueira2019document, nogueira2019doc2query}, using large neural language models, can also be classified as part of this family of approaches, assuming the subsequent retrieval step employs a traditional QTI model like BM25.
In all of these approaches, the focus of the machine learned function is to estimate the impact score of the document with respect to individual terms in the vocabulary, which can be precomputed offline during index creation.

An obvious alternative to document expansion based methods is to use the neural model to reformulate the query~\citep{nogueira2017task, van2017reply, ma2020zero}---although these approaches have not yet demonstrated retrieval performance that can be considered competitive to other methods considered here.

Finally, when the relevance of items are known, or a reliable proxy metric exists, machine learned policies~\citep{kraska2018case, oosterhuis2018potential, rosset2018optimizing} can also be effective for efficient search over indexes but these methods are not directly relevant to our current discussion.

\section{Conformer-Kernel with QTI}
\label{sec:model}

\begin{figure}
\center
\begin{subfigure}{\textwidth}
    \includegraphics[width=\textwidth]{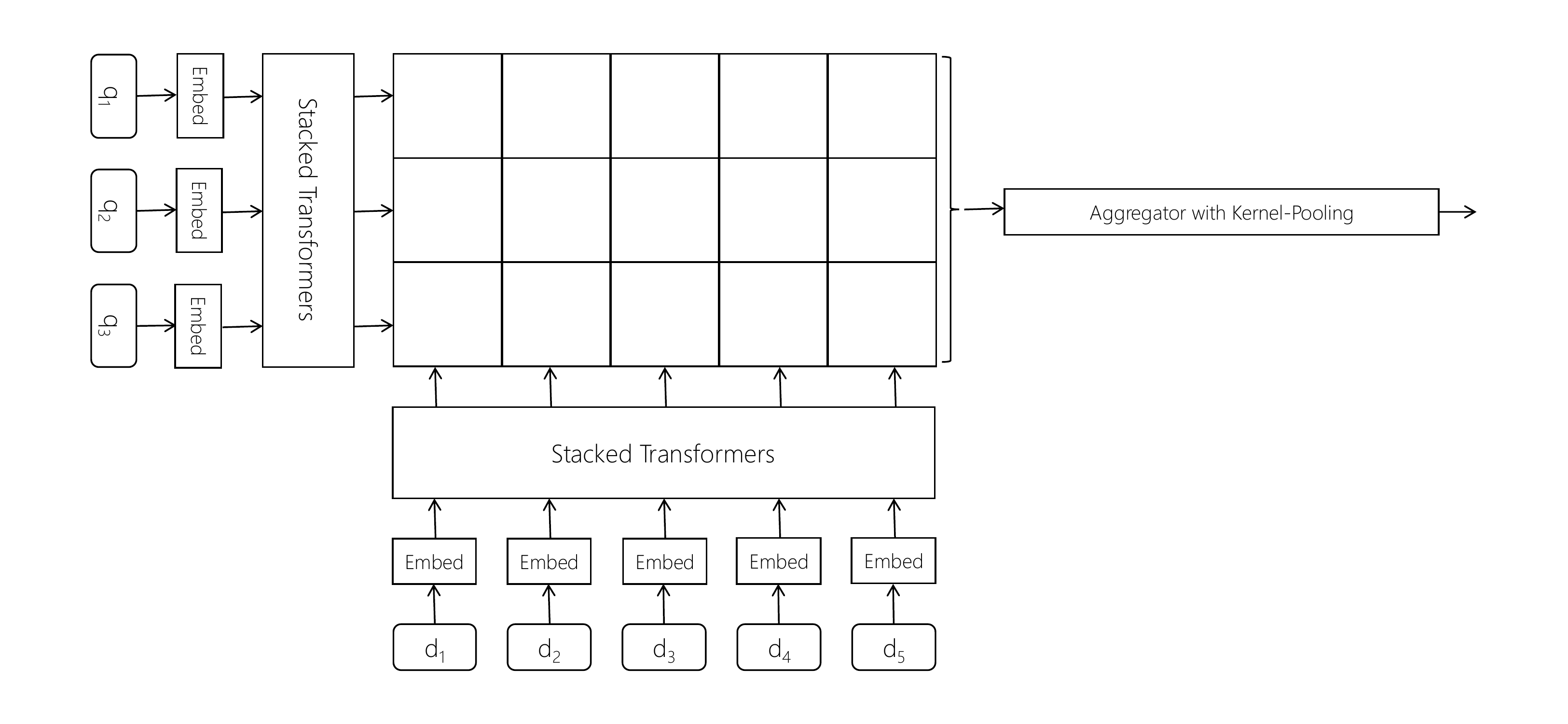}
    \caption{Transformer-Kernel (TK)}
    \label{fig:model-tk}
\end{subfigure}
\hfill
\begin{subfigure}{\textwidth}
    \includegraphics[width=\textwidth]{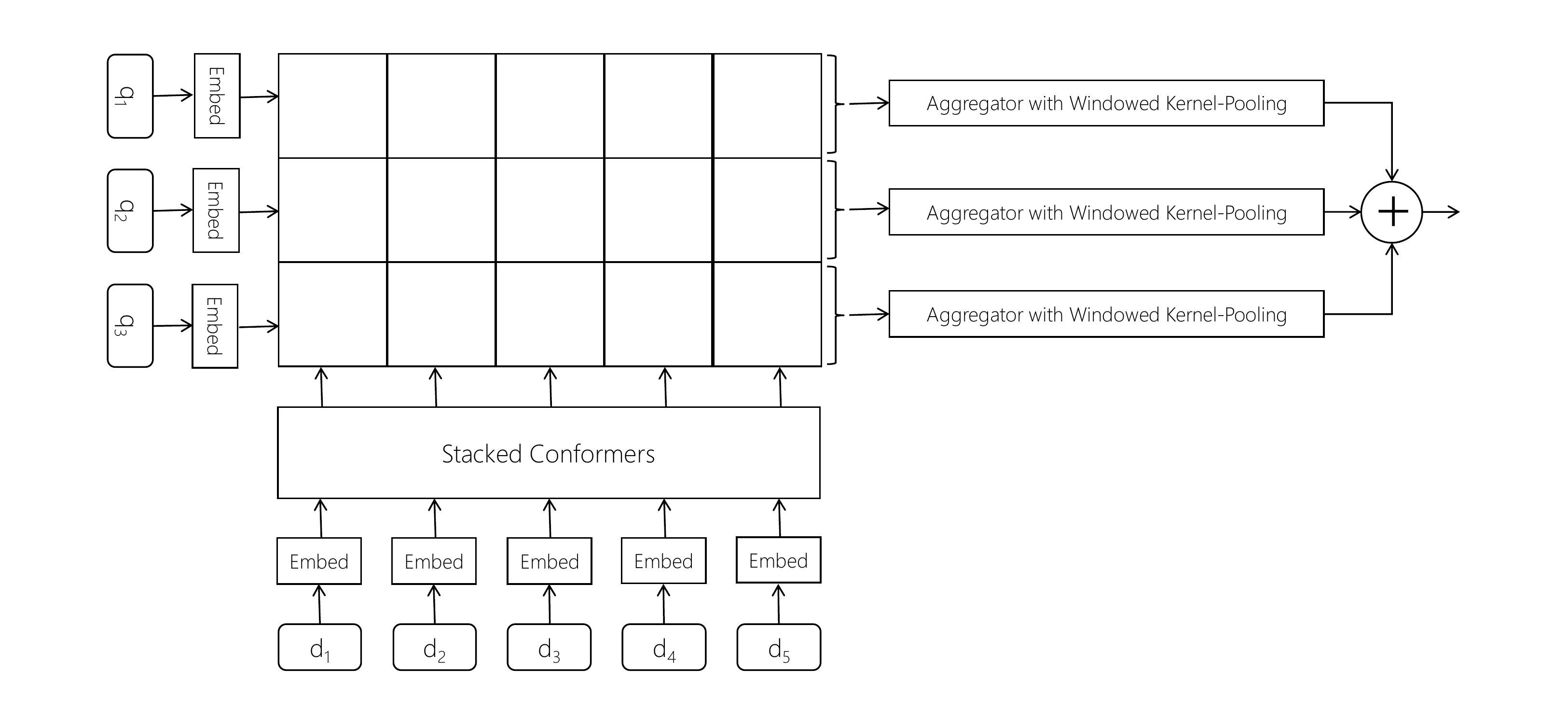}
    \caption{Conformer-Kernel (CK) with QTI}
    \label{fig:model-ck}
\end{subfigure}
\caption{A comparison of the TK and the proposed CK-with-QTI architectures.
In addition to replacing the Transformer layers with Conformers, the latter also simplifies the query encoding to non-contextualized term embedding lookup and incorporates a windowed Kernel-Pooling based aggregation that is employed independently per query term.
}
\label{fig:model}
\end{figure}

We begin by briefly describing the original TK model as outlined in Fig~\ref{fig:model-tk}.
The initial word embedding layer in TK maps both query and document to their respective sequences of term embeddings.
These sequences are then passed through one or more stacked Transformer layers to derive contextualized vector representations for query and document terms.
The learnable parameters of both query and document encoders are shared---which includes the initial term embeddings as well as the Transformer layers.
Based on the contextualized term embeddings, TK creates an interaction matrix $X$, such that $X_{ij}$ is the cosine similarity between the contextualized embeddings of the $i^\text{th}$ query term $q_i$ and the $j^\text{th}$ document term $d_j$.

\begin{align}
    X_{ij} &= \text{cos}(\vec{v_{q_i}}, \vec{v_{d_j}})
\end{align}

The Kernel-Pooling stage then creates $k$ distinct features per query term as follows:

\begin{align}
    K_{ik} &= \text{log}\sum_{j}^{|d|}{\exp{(-\frac{(X_{ij} - \mu_k)^2}{2\sigma^2})}}
\end{align}

Finally, the query-document relevance is estimated by a nonlinear function---typically implemented as stacked feedforward layers---over these features.
Next, we describe the proposed changes to this base architecture.

\subsection{Conformer}
\label{sec:model-conformer}
In Section~\ref{sec:related-long}, we note that the quadratic memory complexity of the self-attention layers \wrt the length of the input sequence is a direct result of explicitly computing the attention matrix $QK^\intercal \in \mathbb{R}^{n \times n}$.
In this work, we propose a new separable self-attention layer that allows us to avoid instantiating the full term-term attention matrix as follows:

\begin{align}
    \text{Separable-Self-Attention}(Q, K, V) &= \Phi(Q) \cdot A \\
    \text{where,} \; A &= \Phi(K^\intercal) \cdot V
\end{align}
As previously, $\Phi$ denotes the softmax operation along the last dimension of the input matrix.
Note that, however, in this separable self-attention mechanism, the softmax operation is employed twice:
\begin{enumerate*}[label=(\roman*)]
    \item $\Phi(Q)$ computes the softmax along the $d_\text{key}$ dimension, and
    \item $\Phi(K^\intercal)$ computes the softmax along the $n$ dimension.
\end{enumerate*}
By computing $A \in \mathbb{R}^{d_\text{key} \times d_\text{value}}$ first, we avoid explicitly computing the full term-term attention matrix.
The memory complexity of the separable self-attention layer is $\mathcal{O}(n \times d_\text{key})$, which is a significant improvement when $d_\text{key} \ll n$.

We modify the standard Transformer block as follows:
\begin{enumerate}
    \item We replace the standard self-attention layer with the more memory efficient separable self-attention layer.
    \item Furthermore, we apply grouped convolution before the separable self-attention layers to better capture the local context based on the window of neighbouring terms.
\end{enumerate}
We refer to this combination of grouped \underline{con}volution and Trans\underline{former} with separable self-attention as a Conformer.
We incorporate the Conformer layer into TK as a direct replacement for the existing Transformer layers and name the new architecture as a Conformer-Kernel (CK) model.
In relation to handling long input sequences, we also replace the standard Kernel-Pooling with windowed Kernel-Pooling~\citep{hofstatter2020improving} in our proposed architecture.

\subsection{Query term independence assumption}
\label{sec:model-qti}
To incorporate the QTI assumption into TK, we make a couple of simple modifications to the original architecture.
Firstly, we simplify the query encoder by getting rid of all the Transformer layers and only considering the non-contextualized embeddings for the query terms. Secondly, instead of applying the aggregation function over the full interaction matrix, we apply it to each row of the matrix individually, which corresponds to individual query terms.
The scalar outputs from the aggregation function are linearly combined to produce the final query-document score.
These proposed changes are shown in Fig~\ref{fig:model-ck}.

\subsection{Explicit term matching}
\label{sec:model-duet}
We adopt the Duet~\citep{nanni2017benchmark, mitra2019updated, mitra2019duet} framework wherein the term-document score is a linear combination of outputs from a latent and and an explicit matching models.

\begin{align}
    s_{t,d} = w_1 \cdot \text{BN}(s_{t,d}^\text{(latent)}) + w_2 \cdot \text{BN}(s_{t,d}^\text{(explicit)}) + b
\end{align}

Where, $\{w_1, w_2, b\}$ are learnable parameters and BN denotes the BatchNorm operation~\citep{ioffe2015batch}.

\begin{align}
    \text{BN}(x) &= \frac{x - \mathbb{E}[x]}{\sqrt{\text{Var}[x]}} 
\end{align}

We employ the CK model to compute $s_{t,d}^\text{(latent)}$ and define a new lexical matching function modeled on BM25 for $s_{t,d}^\text{(explicit)}$.

\begin{align}
    s_{t,d}^\text{(explicit)} &= \text{IDF}_{t} \cdot \frac{\text{BS}(\text{TF}_{t,d})}{\text{BS}(\text{TF}_{t,d}) + \text{ReLU}(w_\text{dlen} \cdot \text{BS}(|d|) + b_\text{dlen}) + \epsilon}
\end{align}

Where, $\text{IDF}_{t}$, $\text{TF}_{t,d}$, and $|d|$ denote the inverse-document frequency of the term $t$, the term-frequency of $t$ in document $d$, and the length of the document, respectively.
The $w_\text{dlen}$ and $b_\text{dlen}$ are the only two leanrable parameters of this submodel and $\epsilon$ is a small constant added to prevent a divide-by-zero error.
The BatchScale (BS) operation is defined as follows:

\begin{align}
    \text{BS}(x) &= \frac{x}{\mathbb{E}[x] + \epsilon} 
\end{align}

\section{Experiments}
\label{sec:experiment}

\subsection{Task and data}
\label{sec:experiment-data}
We conduct preliminary experiments on the document retrieval benchmark provided as part of the TREC Deep Learning track~\citep{craswell2019overview}.
The benchmark is based on the MS MARCO dataset~\citep{bajaj2016ms} and provides a collection of $3,213,835$ documents and a training dataset with $384,597$ positively labeled query-document pairs.
Recently, the benchmark also made available a click log dataset, called ORCAS~\citep{craswell2020orcas}, that can be employed as an additional document description field.
We refer the reader to the track website\footnote{\url{https://microsoft.github.io/TREC-2020-Deep-Learning/}} for further details about the benchmark.

Because we are interested in the full ranking setting, we do not make use of the provided document candidates and instead use the proposed model to retrieve from the full collection.
We compare different runs based on following three metrics: mean reciprocal rank (MRR)~\citep{craswell2009mean}, normalized discounted cumulative gain (NDCG)~\citep{jarvelin2002cumulated}, and normalized cumulative gain (NCG)~\citep{rosset2018optimizing}.

\subsection{Model training}
We consider the first $20$ terms for every query and the first $4000$ terms for every document.
When incorporating the ORCAS data as an additional document field, we limit the maximum length of the field to $2000$ terms.
We pretrain the word embeddings using the word2vec~\citep{mikolov2013efficient, mikolov2013distributed, mikolov2013linguistic} implementation in FastText~\citep{joulin2016bag}.
We use a concatenation of the IN and OUT embeddings~\citep{nalisnick2016improving, mitra2016desm} from word2vec to initialize the embedding layer parameters.
The document encoder uses 2 Conformer layers and we set all the hidden layer sizes to $256$.
We set the window size for the grouped convolution layers to $31$ and the number of groups to $32$.
Correspondingly, we also set the number of attention heads to $32$.
We set the number of kernels $k$ to $10$.
For windowed Kernel-Pooling, we set the window size to $300$ and the stride to $100$.
Finally, we set the dropout rate to $0.2$.
For further details, please refer to the publicly released model implementation in PyTorch.\footnote{\url{https://github.com/bmitra-msft/TREC-Deep-Learning-Quick-Start}}
All models are trained on four Tesla P100 GPUs, with 16 GB memory each, using data parallelism.

We train the model using the RankNet objective~\citep{burges2005learning}.
For every positively labeled query-document pair in the training data, we randomly sample one negative document from the provided top $100$ candidates corresponding to the query and two negative documents from the full collection.
In addition to making pairs between the positively labeled document and the three negative documents, we also create pairs between the negative document sampled from the top $100$ candidates and those sampled from the full collection, treating the former as more relevant.
This can be interpreted as incorporating a form of weak supervision~\citep{dehghani2017neural} as the top candidates were previously generated using a traditional IR function.

\section{Results}
\label{sec:result}

\begin{table}
    \small
    \centering
    \caption{Full retrieval results based on the TREC 2019 Deep Learning track test set.}
    \begin{tabular}{lcccc}
    \hline
    \hline
        \textbf{Model} & \textbf{MRR} & \textbf{NDCG@10} & \textbf{NCG@100} \\
        \hline
        \textbf{Non-neural baselines} \\
        BM25+RM3 run with best NDCG@10 & $0.807$ & $0.549$ & $0.559$ \\
        Non-neural run with best NDCG@10 & $0.872$ & $0.561$ & $0.560$ \\
        \hline
        \textbf{Neural baselines} \\
        DeepCT run with best NDCG@10 & $0.872$ & $0.554$ & $0.498$ \\
        BERT-based document expansion + reranking run with best NCG@10 & $0.899$ & $0.646$ & $0.637$ \\
        BERT-based document expansion + reranking run with best NDCG@10  & $0.961$ & $0.726$ & $0.580$ \\
        \hline
        \textbf{Our models} \\
        Conformer-Kernel & $0.845$ & $0.554$ & $0.464$ \\
        Conformer-Kernel + learned BM25 & $0.906$ & $0.603$ & $0.533$ \\
        Conformer-Kernel + learned BM25 + ORCAS field & $0.898$ & $0.620$ & $0.547$ \\
        \hline
        \hline
    \end{tabular}
    \label{tbl:results}
\end{table}

Table~\ref{tbl:results} presents our main experiment results.
As specified earlier, we evaluate our models on the full ranking setting without any explicit reranking step.
The full model---with both Conformer-Kernel and explicit matching submodel---performs significantly better on NDCG@10 and MRR compared to the best traditional runs from the 2019 edition of the track.
The model also outperforms the DeepCT baseline which is a QTI-based baseline using BERT.
The other BERT-based baselines outperform our model by significant margins.
We believe this observation should motivate future exploration on how to incorporate pretraining in the Conformer-Kernel model.
Finally, we also notice improvements from incorporating the ORCAS data as an additional document descriptor field.

\begin{figure}
\center
\includegraphics[width=0.7\textwidth]{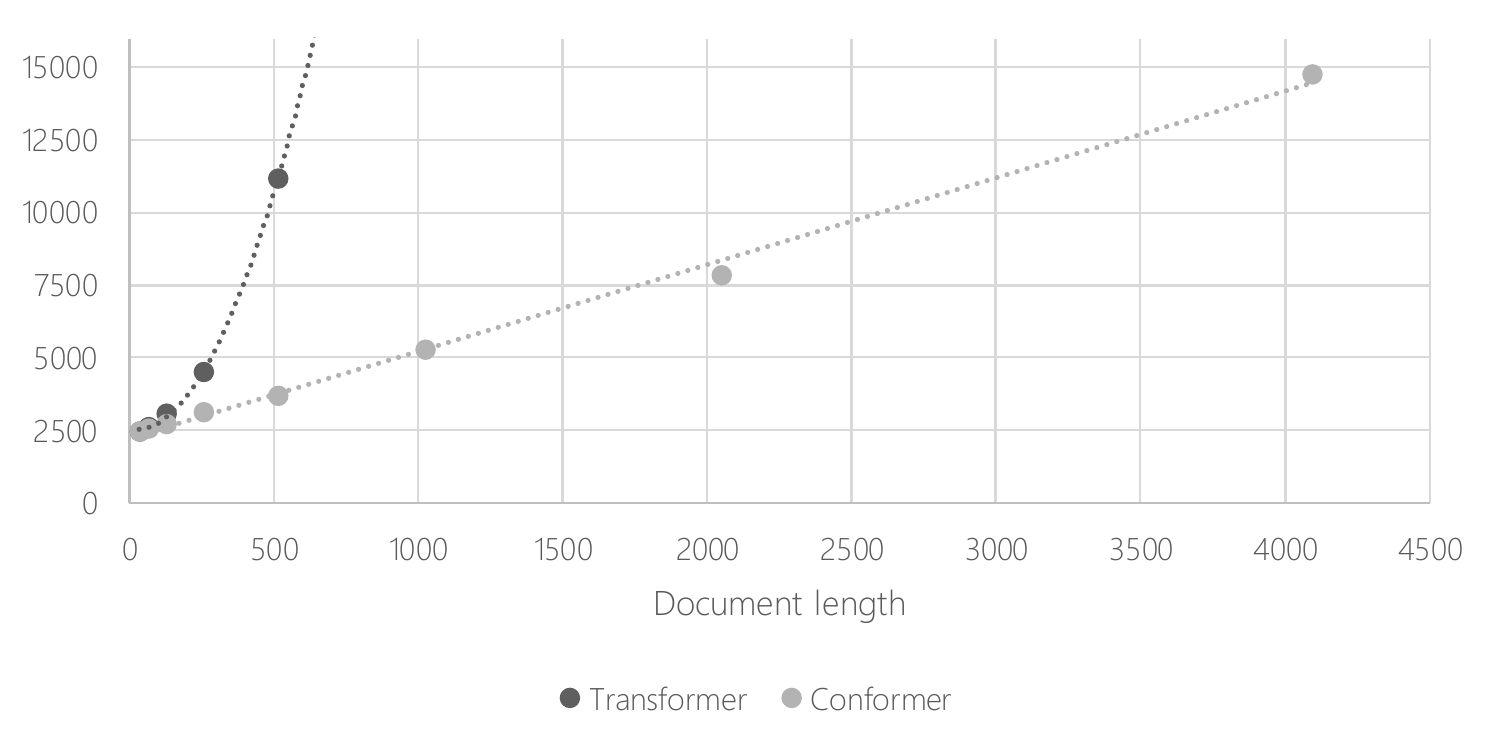}
\caption{Comparison of peak GPU Memory Usage in MB, across all four GPUs, when employing Transformers \vs Conformers in our proposed architecture.
For the Transformer-based model, we only plot till sequence length of 512, because for longer sequences we run out of GPU memory when using Tesla P100s with 16 GB of memory.}
\label{fig:memory}
\end{figure}

To demonstrate how the GPU memory consumption scales with respect to input sequence length, we plot the peak memory, across all four GPUs, for our proposed architecture using Transformer and Conformer layers, respectively, keeping all other hyperparameters and architecture choices fixed.
Fig~\ref{fig:memory} shows the GPU memory requirement grows linearly with increasing sequence length for the Conformer, while quadratically when Transformer layers are employed.
\section{Discussion and future work}
\label{sec:conclusion}

The proposed CK-with-QTI architecture provides several advantages, with respect to inference cost, compared to its BERT-based peers.
In addition to a shallower model and more memory-efficient Conformer layers, the model allows for offline pre-encoding of documents during indexing.
It is notable, that the document encoder, containing the stacked Conformer layers, is the computationally costliest part of the model.
In the proposed architecture, the document encoder needs to be evaluated only once per every document in the collection.
This is in contrast to once per every query-document pair in the case of BERT-based ranking models that accepts a concatenation of query and document as input~\citep{nogueira2019passage}, and once per every term-document pair in the case of BERT-based ranking models with QTI~\citep{mitra2019incorporating}.

While the present study demonstrates promising progress towards using TK-style architectures for retrieval from the full collection, it is worthwhile to highlight several challenges that needs further explorations.
More in depth analysis of the distribution of term-document scores is necessary which may divulge further insights about how sparsity properties and discretization can be enforced for practical operationlization of these models.
Large scale pretraining in the the context of these models also presents itself as an important direction for future studies.
Finally, for the full retrieval setting, identifying appropriate negative document sampling strategies during training poses as an important challenge that can strongly help or curtail the success these models achieve on these tasks.

In the first year of the TREC Deep Learning track, there was a stronger focus on the reranking setting---although some submissions explored document expansion and other QTI-based strategies.
We anticipate that in the 2020 edition of the track, we will observe more submissions using neural methods for the full retrieval setting, which may further improve the reusability of the TREC benchmark~\citep{yilmaz2020reliability} for comparing these emerging family of approaches, and provide additional insights for our line of exploration.

\bibliographystyle{plainnat}
\bibliography{bibtex}

\end{document}